# MegaM@Rt2 EU Project: Open Source Tools for Mega-Modelling at Runtime of CPSs

Jesus Gorroñogoitia Cruz[1], Andrey Sadovykh[2], Dragos Truscan[3], Hugo Bruneliere[4], Pierluigi Pierini[5], Lara Lopez Muñiz[1]

[1]ATOS, Spain
{jesus.gorronogoitia,lara.lopez}@atos.net

[2] Innopolis University, Russia
a.sadovykh@innopolis.ru

[3] Åbo Akademi University, Finland
dragos.truscan@abo.fi

[4] IMT Atlantique, LS2N (CNRS) & ARMINES, Nantes, France
hugo.bruneliere@imt-atlantique.fr

[5] Intecs Solutions S.p.A., Italy
pierluigi.pierini@intecs.it

**Abstract.** In this paper, we overview our experiences of developing large set of open source tools in ECSEL JU European project called MegaM@Rt2 whose main objective is to propose a scalable model-based framework incorporating methods and tools for the continuous development and runtime support of complex software-intensive Cyber-Physical Systems (CPSs). We briefly present the MegaM@Rt2 concepts, discuss our approach for open source, enumerate tools and give an example of a tools selection for a specific industrial context. Our goal is to introduce the reader with open source tools for the model-based engineering of CPSs suitable for diverse industrial applications.

## 1   Introduction

MegaM@Rt2 is a three-years project, which started in April 2017[1], [2] and which is funded by European Components and Systems for European Leadership Joint Undertaking (ECSEL JU) under the H2020 European program.  The main goal of  MegaM@Rt2 is to create an integrated framework incorporating scalable methods and tools for continuous system engineering and runtime validation and verification (V&V). The framework addresses the needs of the 8 case study providers involved in the project, which come from diverse and heterogeneous industrial domains, ranging from transportation and telecommunications to logistics and manufacturing. The underlying objective is to provide improved productivity, quality, and predictability of large and complex industrial cyber-physical systems (CPSs).

In order to address these needs, 20 technology and research providers contributed with more than 28 tools integrated into the MegaM@Rt2 toolbox. The results enumerated in this paper are a complement to the research achievements discussed in [3]. In this paper, we briefly present the MegaM@Rt2 concept, discuss the choice for the open source in the project, enumerate the open-source tools provided by the project, and finally give an example of an open source tool chain for a telecom application.

## 2   MegaM@Rt2 Overall Concept

The project addresses the fundamental challenge to support efficient forward and backward traceability between the two main system levels: design-time and runtime. In parallel to these, modern large-scale industrial software engineering processes require thorough configuration and model governance to provide the promised productivity gains. As an answer to the above challenge, MegaM@Rt2 provides a scalable mega-modelling approach to manage all the involved artifacts including the multitude of different types of models, corresponding workflows, and configurations, among others. In this context, an important challenge is to better tackle large diversity of models in terms of nature, number, size, and complexity.

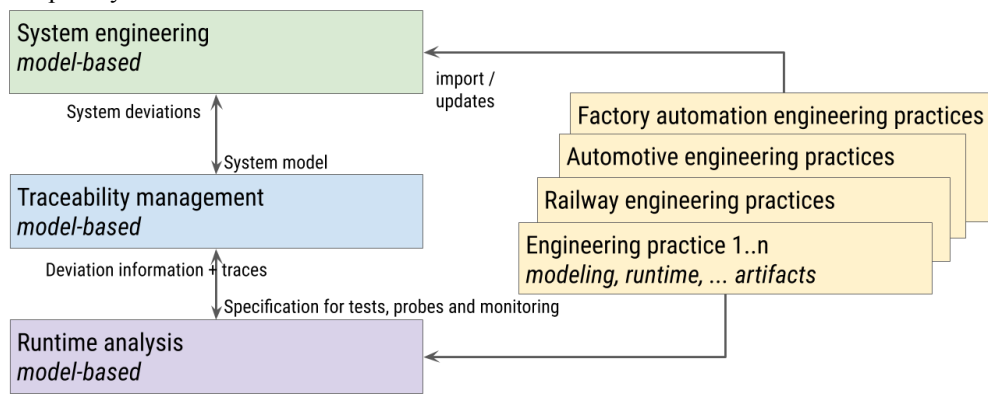

**Fig. 1.** Conceptual architecture of the MegaM@Rt2 project

Conceptually (Fig. 1) the project proposes to integrate various system modelling artifacts on the System engineering level, while the Runtime artifacts such as tests and on-line monitors produce traces that have to be analysed and linked with system artifacts for validation and remediation purposes. Consequently the project provides numerous tools that are categorised in those above-mentioned areas: *a) system engineering*, *b) runtime analysis, and c) traceability management*. The end users, industrial case studies, select a subset of the MegaM@Rt2 tools based on their preferred methodologies and technical areas. For example in the railway domain, engineers follow V-lifecycle and focus on safety with specific modelling and verification techniques. They benefit from the large variety of the analysis tools that are enabled with the common traceability mechanisms.

Open source is important for the project in many aspects. Releasing tools as open source is important in the project in order to increase the audience and adoption of the tools by the community and industrial partners. Moreover, the project promotes wide adoption of its methods through open source along with other methods. Finally, open source approach is a mechanism to ensure sustainability aiming to create a community of interest around project results and ensure industry for the perennity of their preferred tools.

## 3   MegaM@Rt2 Contribution to Open Source

In order to put in practice the global approach presented in the previous section, this section introduces the MegaM@Rt2 toolbox. This toolbox is composed of three complementary tool sets covering System Engineering, Runtime Analysis, and Model & Traceability Management, respectively. The toolbox is freely accessible from the MegaM@Rt2 web portal [4] offers detailed information about all the available MegaM@Rt2 tools, including links to their main artifacts such as software downloads, documentation and source code, available in a public repository. In addition, this web portal provides capabilities that facilitate the searching of suitable tools by keyword (i.e., tag cloud) and license type. In this paper, we focus on the tools provided under an open source license scheme and how they have been managed and maintained in the project. The use of open source distribution licenses facilitates a wider adoption of the tools and their support, maintenance and evolution by their own community of users.

Most of the open source MegaM@Rt2 tools are Eclipse-based, which can be installed, as plugins, within an existing Eclipse 2018-09 version. This version has been adopted as the baseline for all tools to be made compatible with, in order to facilitate their integration within a common Eclipse framework. Besides, the MegM@Rt2 toolset packaging and delivery approach is based on common Eclipse packaging and delivery mechanisms: i) a public MegaM@Rt2 Eclipse update site [5] that users can apply to select and install the MegaM@Rt2 tools, and ii) the MegaM@Rt2 IDE (including tools and dependencies) has been published in the Eclipse Marketplace [6], from where users can install it. These mechanisms are well-known among the Eclipse community and they largely simplify the installation of the MegaM@Rt2 toolset. Furthermore, downloadable bundles [7] of the MegaM@Rt2 IDE toolsets with all required dependencies installed have been packaged for Windows, MacOS and Linux.

The tools included in the latest version of the Eclipse bundle can be distributed according to the work packages in the project as follows: a) *system engineering* - Papyrus and Moka extensions for aspect-oriented modeling and fUML simulation logging, respectively, Collaboro for collaborative language development, EMFToCSP for automatic model verification, S3D for designing the software and hardware of embedded systems, HepsyCode for HW/SW Co-Design of Heterogeneous Parallel Dedicated Systems; b) *runtime analysis* - PADRE for model refactoring, VeriATL for model transformation verification; and c) *traceability and megamodeling* - NeoEMF for scalable model loading and handling, EMF Views for building model views and JTL for traceability management.

Additional tools contributing to system engineering are distributed as Eclipse Rich Client Platform applications, e.g. CHESS [8] - for developing hard real-time, safety-critical and high-integrity embedded systems, or can be downloaded from their developer web site, e.g.

Modelio [9] for system modeling in UML, SysML and MARTE; and Refinement Calculus of Reactive Systems - RCSR for model verification and reasoning.

Other MegaM@Rt2 open source tools that are not Eclipse-based are packaged (together with their dependencies and testing examples) and delivered within containerized packages, which can be generated from downloadable images. MegaM@Rt2 offers Docker images [10] and Linux scripts to build tool containers and run the tools right from those images. This approach largely simplifies the burden of installing the tools and their dependencies, and does not require any knowledge of the Docker technology from end-users. For instance, MegaM@Rt2 provides docker images for system engineering, such as PauWARE - for instrumenting Statecharts execution in Java and for run time analysis, such as AIPHS - for on-chip monitoring, and for runtime analysis, such as the LIME tool - for testing and runtime monitoring.

In terms of licensing schemes, the above mentioned tools are released under one of the open source licenses: copyleft or viral licenses, such as GPLv2/3 or LGPLv3, semi-restrictive licenses such as EPLv1/v2, and permissive licenses such as APLv2, and MIT. The choice of licensing scheme for each tool was dictated both by the dependencies of each tool and by the interest of the tool vendors.

A detailed description of the capabilities of each tool in the MegaM@Rt2 toolbox, including how it satisfies the requirements of the industrial case studies in the project, can be found in public project deliverables covering the three tool sets previously mentioned D2.5, D3.5 and D4.4, respectively, released in 2019 [11]. Moreover, a second set of public project deliverables provide conceptual aspects, methodologies, and guidelines on how each tool in the toolbox can be used by end users D2.6, D3.6 and D4.5 [11].

## 4    Open Source Toolchain for a Telecom Case Study

The Teknè case study is a concrete example of how open source tools belonging to the MegaM@Rt framework are applied to solve industrial needs and how MegaM@Rt tool set allows easy and flexible integration with external tools. Actually, the last point is a general issue in the industrial practice where internal processes require formal procedures and consolidated tools. MegaM@Rt addressed the problem by providing a framework that can be easily tailored to user needs, e.g. selecting the subset of needed tools, and integrated in external context exploiting model transformation techniques.

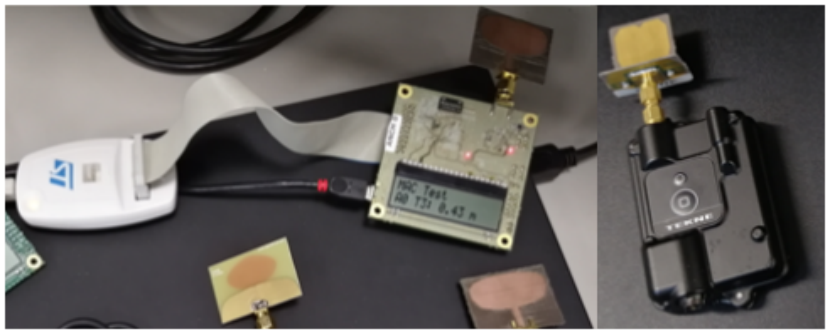

**Fig. 2.** Tekne wearable mote example.

The case study is centered on a *wearable mote* (see Fig. 2), which is a mobile network node, based on the Ultra-Wideband (UWB) technology, with short range communications, indoor positioning and tracking capabilities. It can be used, for instance, to evaluate collision risks among a set of mobile devices in a construction site.

Among the tools adopted to solve this case study, CHESS, used for design and analysis, and ρEmbedded libraries, used to instrument code with monitor probes required to collect runtime traces, are open source tools part of the MegaM@Rt framework, while Yakindu [12] is a commercial tool used for code generation, but a free license is available for non-commercial and academic use.

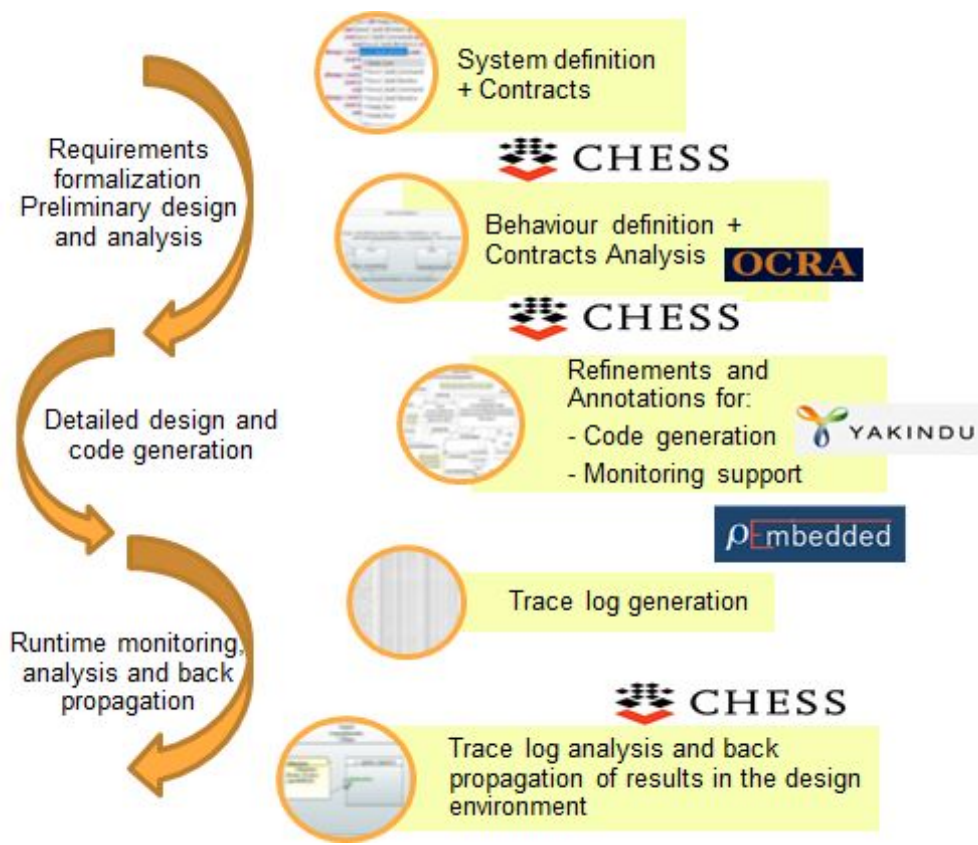

**Fig. 3.** Tekne case study tool chain and process.

The CS implementation process shown in Fig. 3 is split in three main steps: i) the requirements analysis and formalisation, the preliminary design and analysis; ii) detailed design and code generation, iii) log collection at runtime and log analysis and back propagation of the results to the design step. In the first step, the CHESS contract-based design approach has been exploited. The system requirements are analysed to derive the system architecture and then they are formalised, deriving the "assumption/guarantee" pairs, using Linear Temporal Logic (LTL), to compose contracts. Contracts are associated with system architectural components. CHESS is seamlessly integrated with the OCRA tool [13] supporting the analysis of the components behavior (expressed as a set of finite state machines) with respect to the associated contracts, to assess the formal correctness of the architecture.

During the second step, the implementation components are derived from the architectural model and MARTE annotations are added to mark the timing constraints to be checked at runtime. The resulting model is exported to Yakindu that allows additional annotation for code generation and instrumentation with monitoring support based on ρEmbedded libraries.

The final step provides the collection of the log traces; logs are then given in input to CHESS for the analysis of the non-functional characteristics of the system at runtime and the back propagation of the results to the design environment. In particular the derived properties are traced back to the relevant implementation components of the CHESS model to allow verification with respect to the defined constraints/requirements. Further refinements of the design are then applied if necessary.

The contract based approach supported by CHESS enables the formalization of functional and non functional requirements, and the early validation of the model, while runtime monitoring capabilities provides effective validation tests of the performances that are back annotated in the design model. The approach provides a strong association (i. e. traceability) among requirements, contracts, architectural components and performance values measured at runtime.

The evaluation of the benefits obtained by MegaM@Rt2 focus mainly on design efforts (both in terms of resources costs and development time) and the quality of the final results. The exploitation of the runtime artefacts backtraced to the design models and to original requirements, greatly simplifies the maintenance activities and provides a mechanism to continuously improve design, increasing the level of the obtained benefits.

The use of open source tools foster the experimentation at industrial level of innovative and advanced technologies, highlighting both the positive economic impacts, as mentioned above and the problems related to the complexity of real systems, such as scalability, stability, and ease to use of the open source research solutions. This allowed technology providers to raise the tools proposed in the MegaM@Rt2 framework to an industrial scale.

## 5  Conclusions

Development of complex CPSs is a challenging activity requiring a combination of many tools. Open source tools for model-based system engineering have gained certain popularity. However, there is still a challenge to combine them into a practical tool chain that would address specific application domain needs. The tool chain should ensure interplay of diverse tools, as well as a transversal traceability for model artifacts. This is especially challenging in the modern context when the runtime analysis of target systems should be analysed with the help of system engineering models for verification and validation. The MegaM@Rt2 project focuses on providing tool sets that deal with model-based system engineering, runtime analysis, and traceability management. MegaM@Rt2 consortium is an active contributor to open source and has extensively applied open source tools within 8 diverse industrial case studies. In this paper, we have presented the overall concept of MegaM@Rt2, we have given indications on open source tools provided for various engineering areas and on the approaches for distribution of a large bundle of open source tools, and, finally, we have illustrated the benefits of our approach with a practical industrial case study example. We believe that this paper is valuable for the community of the CPS developers looking for an operational open source tool set for their purposes.

## Acknowledgments

This work has received funding from the Electronic Component Systems for European Leadership Joint Undertaking under grant agreement No 737494. This Joint Undertaking receives support from the European Union's Horizon 2020 research and innovation programme and Sweden, France, Spain, Italy, Finland, the Czech Republic.